\documentclass[aps,twocolumn,showpacs,preprintnumbers,amsmath,amssymb,superscriptaddress]{revtex4-1}
\usepackage{graphicx}
\usepackage{dcolumn}
\usepackage{bm}
\usepackage{color}
\usepackage{url}

\newcolumntype{d}[1]{D{.}{\cdot}{#1}}
\newcolumntype{.}{D{.}{.}{-1}}
\newcolumntype{,}{D{,}{,}{-1}}
\DeclareTextCompositeCommand{\r}{OT1}{A}{%
  \leavevmode\vbox{%
    \offinterlineskip
    \ialign{\hfil##\hfil\cr\char23\cr\noalign{\kern-1.15ex}A\cr}%
  }%
}	

\begin{document}

\title{Machine-learning Based Extraction of the Short-Range Part of the Interaction in Non-contact Atomic Force Microscopy}
\author{Zhuo Diao}
\affiliation{Graduate School of Engineering Science, Osaka University, 1-3 Machikaneyama, Toyonaka, Osaka 560-0043, Japan}
\author{Daiki Katsube}
\affiliation{Graduate School of Engineering Science, Osaka University, 1-3 Machikaneyama, Toyonaka, Osaka 560-0043, Japan}
\affiliation{Graduate School of Engineering, Nagaoka University of Technology, 1603-1 Kamitomiokamachi, Nagaoka, Niigata 940-2188, Japan.}
\author{Hayato Yamashita}
\affiliation{Graduate School of Engineering Science, Osaka University, 1-3 Machikaneyama, Toyonaka, Osaka 560-0043, Japan}
\author{Yoshiaki Sugimoto}
\affiliation{Department of Advanced Materials Science, The University of Tokyo, Kashiwanoha 5-1-5, Kashiwa, Chiba 227-8561, Japan}
\author{Oscar Custance}
\affiliation{National Institute for Materials Science (NIMS), 1-2-1 Sengen, Tsukuba, Ibaraki, Japan}
\author{Masayuki Abe}
\email{abe@stec.es.osaka-u.ac.jp}
\affiliation{Graduate School of Engineering Science, Osaka University, 1-3 Machikaneyama, Toyonaka, Osaka 560-0043, Japan}

\date{\today}

\begin{abstract}
A machine-learning method for extracting the short-range part of the probe-surface interaction from force spectroscopy curves is presented. 
Our machine-learning algorithm consists of two stages: the first stage determines a boundary that separates the region where the short-range interaction is dominantly acting on the probe, and a second stage that finds the parameters to fit the interaction over the long-range region. 
We successfully applied this method to force spectroscopy maps acquired over the $\mathrm{Si}(111)-(7\times7)$ surface and found, as a result, a faint structure on the short-range interaction for one of the probes used in the experiments that would have probably been obviated using human-supervised fitting strategies. 
\end{abstract}

\maketitle



Machine-learning techniques have been used in fields such as robotics\cite{Hwangboeaau5872, Gu:2016aa}, computer vision\cite{Thrun:2003:RMS:779343.779345, 1802.08195}, natural language processing\cite{1409.0473}, and games\cite{Silver:2016aa}, among others, to train machines to perform experience-based tasks in a smart way.
Recent breakthroughs on machine-learning enable researchers to automatically analyze a big amount of data and interpret the results in a better way.
Successful results have been obtained in the fields of geoscience\cite{Bergeneaau0323}, genome\cite{Zhang:2017aa},  medicine\cite{Gurovich:2019aa} and material science\cite{Raccuglia:2016aa}, where not only predictions and classifications\cite{Ren:2017aa,  Jong:2016aa, Ward:2016aa} have been achieved, but also machine learning has helped to develop new materials and devices\cite{Bartok:2017aa, Gomez-Bombarelli:2016aa, Jacksoneaav1190}. 
 In the field of scanning probe microscopy (SPM), machine learning techniques have also been applied to automated imaging analysis\cite{Alldritteaay6913, acsnano.8b02208}.

Non-contact atomic force microscopy (NC-AFM) is a SPM technique that plays an important role as a tool for investigating surfaces and that deals with a large amount of data, especially in the case of force spectroscopy mapping\cite{ncafm1, ncafm2, ncafm3}.
In NC-AFM, the cantilever is oscillated at resonance, and the shift of its resonant frequency due to the interaction of the cantilever's probe with the surface is detected\cite{Albrecht1991}.
The dependence of this frequency shift ($\Delta f$) with the probe-surface separation ($\Delta f -z$), known as force spectroscopy, gives us insight into the surface properties. One can calculate distance-dependent force and interaction potential curves from $\Delta f -z$ curves, and even compute lateral forces from two-dimensional $\Delta f -z$ maps\cite{2001Sci...291.2580L, Abe:2005ib, Abe:2007jm, 2008PhRvB..77s5424S, Ternes:2008ka}.
In general, force spectroscopy analysis is carried out under the idea that the probe-surface total interaction includes both short-range and long-range components, with the short-range part being responsible for atomic resolution\cite{PhysRevLett.78.678, Gross1110}.
The long-range component is, in most of the cases, ascribed to van der Waals and electrostatic forces, which are more widespread than the short-range forces and rarely contribute to the atomic contrast.

The extraction of the short-range part of the interaction from force spectroscopy curves is a crucial step to obtain information about single atoms and molecules at surfaces. 
Usually, the long-range part is fitted over a region $z>z_0$ where the short-range interaction is inexistent, and this fit is later on subtracted to the curve to obtain the short-range part. 
Traditionally, this fitting and substation procedure is implemented once the $\Delta f -z$ curve has been converted to the probe-surface interaction force using one of the inversion procedures available in the literature\cite{dvalue, Sader:2004kt}; as for the fitting function, a rational physical model describing a suitable probe-surface interaction force far from the surface is normally used\cite{longshort1}. 
In principle, a similar fitting and substation procedure can also be applied to the $\Delta f -z$ curve to obtain the short-range interaction. 
This is highly desirable when aiming at a full force spectroscopy automation process during measuring, in which the short-range interaction can be evaluated right after the acquisition of a $\Delta f -z$ curve.
Either using the total force or the $\Delta f$ approach, the fitting over the long-range part of the interaction requires an appropriate determination of the threshold distance\cite{longshort1}, $z_0$. 
When there are atomic vacancies \textemdash like in the case of the corner-hole of the $\mathrm{Si}(111)-(7\times7)$ surface\cite{2001Sci...291.2580L}\textemdash\space or nearby areas where there is a lack of short-range interaction\cite{Ternes:2008ka, Sweetman2014}, a curve measured over these specific surface points can be used  to locally characterize the long-range contribution\cite{Abe:2005ib}. 
For a more general case in which there are no atomic vacancies or the surface is not homogeneous in composition or structure, it is challenging to evaluate $z_0$, and automatic methods to obtain this threshold distance are highly desirable.

In this manuscript, we present a machine-learning based scheme  to extract  the short-range part of the interaction from the measured $\Delta f-z$ curves. 
An appropriate selection of the loss functions for the machine-learning algorithm and the introduction of a method for the determination of $z_0$ enable us to evaluate the short-range part of a $\Delta f-z$ curve and to obtain additional contributions to the probe-surface interaction. 
Our analysis of two-dimensional $\Delta f$ maps reveals, for instance, a faint structure on the short-range interaction for one of the probes used in the experiments. 
This faint structure is slightly shifted in position and extends further from the surface than the short-range interaction producing atomic contrast in NC-AFM.


Our machine-learning algorithm consists of two stages: the first stage evaluates $z_0$, and the second stage finds the parameters for the best fit  ($\Delta f_{\mathrm{FIT}}$) over the long-range part of the curve. In  both stages, we use the gradient descent method (GDM)\cite{Baldi1995, DBLP:journals/corr/Ruder16} and update $\Delta \tilde{f}_{\mathrm{FIT}}$, which is a regression model temporal output in each epoch (or iteration) until the loss function in each stage is stabilized within a certain range. 
For simplicity,  we assume a hypothesis function to describe the long-range contribution which is based on a generalized hyperbolic function\cite{longshort1}.
\begin{equation}
\label{formula:fitting}
\Delta f_{\mathrm{FIT}}(z)=a-\frac{b}{(1+cz)^d}, 
\end{equation}
where $a$, $b$, $c$, and $d$ are fitting parameters.

At large enough probe-surface separations, the slope of the short-range component of the $\Delta f-z$ curve should be close to zero, as there is no contribution of the short-range force over the long range interaction region.
Therefore, the value of $z_0$ can be chosen as the distance at which a steep change in the slope of the $\Delta f_{\mathrm{SR}}-z$ curve appears. 
We apply the Otsu method\cite{Otsu1979ieee} to find the threshold in the slope change, and a  $\tilde{z}_0$ value is sequentially updated from the short-range curve $\Delta \tilde{f}_{\mathrm{SR}}=\Delta f -\Delta \tilde{f}_{\mathrm{FIT}}$.
In the first stage of the algorithm, we employ a loss function $E_{\mathrm{1st}}$ using an absolute value error defined as $E_{\mathrm{abs}}(z)= |\Delta \tilde{f}_{\mathrm{FIT}}(z)-\Delta f(z)|$:
\begin{equation}
\label{eq:lossFunction1st}
E_{\mathrm{1st}} =E_{\mathrm{abs}}(z)+w(\tilde{z}_0)E_{\mathrm{abs}}(z>\tilde{z}_0),
\end{equation}
where 
$w(z)$ is a weight function to control the second term contribution to $E_{\mathrm{1st}}$.
The value of $E_{\mathrm{1st}}$ is dynamically updated due to the weight function $w(z) \propto \frac{1}{z_{\mathrm{max}}-z}$ where $z_{\mathrm{max}}$ is the largest separation of the measured $\Delta f$ curve\cite{lossFunction}.
%

In the second stage of the algorithm, we also perform GDM using the measured $\Delta f$ at $z>z_0$ with a loss function of
\begin{equation}
\label{eq:lossFunction2nd}
E_{\mathrm{2nd}} =w_0E_{\mathrm{abs}}(z>z_0),
\end{equation}
where $w_0$ is a constant value.
From a preliminar analysis of the results and the calculation time, a reasonable value for $w_0$ is $\sim 100$.
The frequency shift curve associated to the short-range interaction ($\Delta f_{\mathrm{SR}}-z$) is obtained by the subtraction of $\Delta f_{\mathrm{FIT}}$ to the measured $\Delta f$ ({\it i.e.}, $\Delta f_{\mathrm{SR}}=\Delta f-\Delta f_{\mathrm{FIT}}$).

%
\begin{figure}[tb]
\begin{center}
\includegraphics[width=85mm, clip]{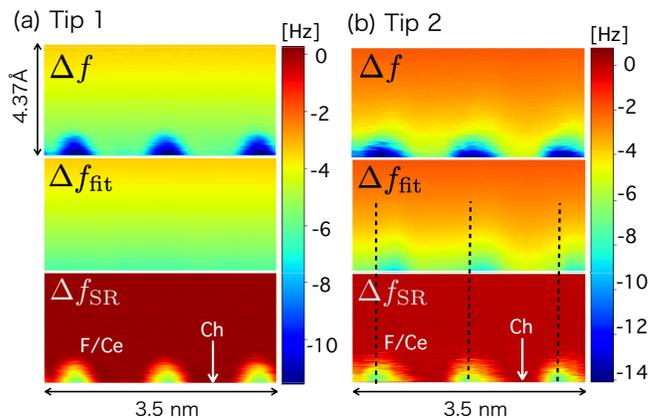}
\caption{\label{fig:SiMapB}
Machine-learning extraction of the short-range interaction from  frequency shift ($\Delta f$) maps measured over the $\mathrm{Si}(111)-(7\times7)$ surface.
Two cantilevers with the same specifications but different probe apexes (Tip 1 and Tip 2) were used for the measurements shown in (a) and (b), respectively.
The $\Delta f_{\mathrm{FIT}}$ maps at the center panels are the output of our machine-learning implementation.
The $\Delta f_{\mathrm{SR}}$ maps at the bottom are calculated by subtracting $\Delta f_{\mathrm{FIT}}$ to the measured $\Delta f$ ({\it i.e.}, $\Delta f_{\mathrm{SR}}=\Delta f-\Delta f_{\mathrm{FIT}}$).
Each of the $\Delta f$ maps consists of 1024 curves acquired along the main diagonal over the faulted-half of the unit cell of the $(7\times7)$ reconstruction.
The symbols ``F/Ce" and ``Ch" highlight the position of the faulted center adatom and the corner hole, respectively.
Only a zoom of the closest $4.37\mathrm{\AA}$ of the total of $21.86\mathrm{\AA}$  explored in the experiment are displayed.
}
\end{center}
\end{figure}

To optimize the GDM implementation, we used the Adam; an algorithm for first-order gradient-based optimization\cite{DBLP:journals/corr/KingmaB14}  that is available in Google TensorFlow\texttrademark\cite{tf}.
We introduce a limit $d \le 1.5$, which is consistent with a van der Waals force dominating the parameter $d$\cite{dvalue}.
This limit enables us to run stable calculations without fluctuations in the fitting parameters.
The data acquisition was performed at room temperature, and the $\Delta f$ maps used in this work were acquired on a $\mathrm{Si(111)-(7\times 7)}$ surface at which a small amount of hydrogen atoms were purposefully adsorbed.
Additional experimental parameters can be found elsewhere\cite{PhysRevB.87.155403}.

%
\begin{figure}[tb]
\begin{center}
\includegraphics[width=85mm, clip]{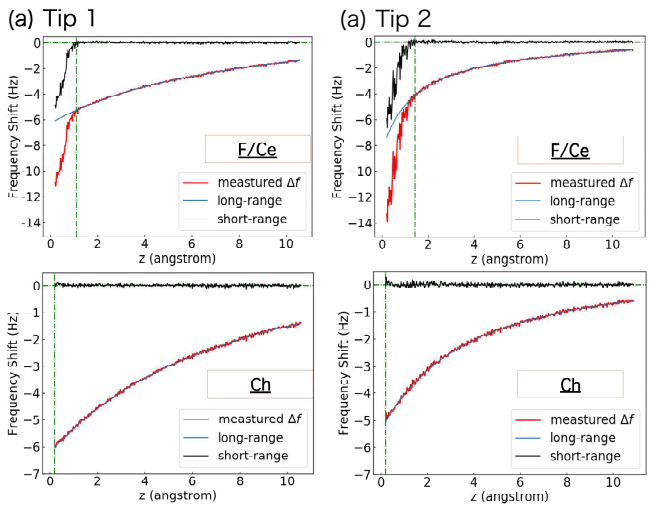}
\caption{\label{fig:FittedCurve}
Line profiles of the $\Delta f$ maps displayed in Fig. \ref{fig:SiMapB}, showing individual $\Delta f-z$ curves measured over the faulted center adatom (F/Ce) and corner hole (Ch) sites for Tip 1 and Tip 2, respectively. The red lines correspond to the measured $\Delta f$, the blue lines are the fitting curves obtained from our machine-learning algorithm, and the black curves correspond to the subtraction of the fitting to the experimental curve. A dash-dotted vertical line highlights the value of $z_0$ calculated in the first stage of our algorithm. 
}
\end{center}
\end{figure}
%

One of the most remarkable properties of our machine-learning method is a batch processing of the short-range part of the interaction in force spectroscopy maps composed of a great number of curves. 
Figure \ref{fig:SiMapB} shows the result of applying our machine-learning implementation to a $\Delta f$ map acquired on the $\mathrm{Si}(111)-(7\times7)$ surface and composed by 1024 curves.
The experimental $\Delta f$ maps shown at the top of Fig. \ref{fig:SiMapB} were acquired with two different AFM tips over a total probe-surface separation of $21.86\r{A}$.
We applied our machine-learning method to the $\Delta f$ maps, and obtained the $\Delta f_{\mathrm{FIT}}$ maps displayed at the middle of Fig. \ref{fig:SiMapB}, with the corresponding parameters.
The $\Delta f_{\mathrm{SR}}$ maps at the bottom are obtained by subtracting $\Delta f_{\mathrm{FIT}}$ to the experimental $\Delta f$ ({\it i.e.}, $\Delta f_{\mathrm{SR}}=\Delta f-\Delta f_{\mathrm{FIT}}$).

Figure \ref{fig:SiMapB} evidences the relevance of measuring two- and three-dimensional maps of $\Delta f$, as well as the usefulness of our machine-learning method for analyzing them.
A striking result is obtained from the $\Delta f_{\mathrm{FIT}}$ maps, which present a different behavior for the two probes used. 
In the case of Tip 1, the $\Delta f_{\mathrm{FIT}}$ values approaches towards zero as the probe-surface separation increases without showing any apparent feature at the probed atomic sites: the $\Delta f_{\mathrm{FIT}}-z$ curves are almost the same at every position of the map, and only the long-range component of the force dominates the probe-surface interaction ($\Delta f_{\mathrm{FIT}} \approx \Delta f_{\mathrm{LR}}$). 
For Tip 2, on the contrary, a site-specific pattern clearly appears in the $\Delta f_{\mathrm{FIT}}$ map. 

In order to verify that the machine-learning based fitting was correctly executed, we compared sets of $\Delta f-z$, $\Delta f_{\mathrm{FIT}}-z$, and $\Delta f_{\mathrm{SR}}-z$ curves acquired over the faulted center adatom (F/Ce) and the corner hole (Ch) sites for both probes, as it is shown in Fig. \ref{fig:FittedCurve}. 
The value of $z_0$ calculated for each site in the first stage of the algorithm is indicated by a vertical dash-dotted line. 
The interaction between the Si atoms of the surface and the atoms at the probe apex produces an abrupt decrease in the $\Delta f$ signal over the adatom site. 
As expected, this abrupt change is missing at the corner holes\cite{Lantz2580}.
For both tips, similar $\Delta f-z$ curves were obtained at the respective sites. 
At the F/Ce sites, the first stage of the calculation produces $z_0 = 1.1\mathrm{\AA}$ for Tip 1 and $z_0 = 1.4\mathrm{\AA}$ for Tip 2; separations at which an abrupt decreases in the $\Delta f$ signal is registered.
At the corner hole, an output of $z_0 \approx 0$ is consistent with the fact that at this site probe-surface short-range forces are negligible at the separations explored in the experiment. 
In Fig. \ref{fig:FittedCurve}, the fitting over the long-range part of the $\Delta f-z$ curve and the result of its subtraction to the experimental $\Delta f$ curves are also shown, validating a correct output of the automated calculation.


The site-specific characteristic of the pattern visible in the $\Delta f_{\mathrm{FIT}}$ map for the case of Tip 2 (Fig. \ref{fig:SiMapB}b) seems to be related to the short-range interaction, as the features span over a distance of $\approx 2\mathrm{\AA}$ from the closest approach to the surface.  
A detailed comparison of these patterns with the features ascribed to the onset of the bonding of the probe-surface closest atoms ($\Delta f_{\mathrm{SR}}$ map) reveals that the features in the $\Delta f_{\mathrm{FIT}}$ map are slightly shifted to the right side by approximately $2.2\mathrm{\AA}$. These results suggest that an additional component to the short-range force could be acting in the case of Tip 2.

\begin{figure}[tb]
\begin{center}
\includegraphics[width=85mm, clip]{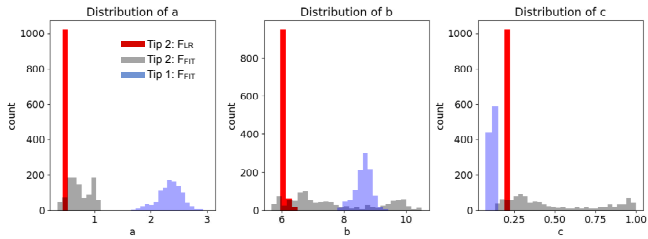}
\caption{\label{fig:histogram}
Histograms of the parameters $a, b$ and $c$ of our fitting model obtained by our machine-learning implementation.
The distributions of parameters that are obtained running the first and second stages of the calculation are shown in purple and grey for Tip 1 and 2, respectively.
The histograms in red represent the distribution of parameters for Tip 2 when skipping the first stage and carrying out the second stage preselecting $z_0=7.47\mathrm{\AA}$ and $d =1.5$. 
}
\end{center}
\end{figure}

\begin{figure}[tb]
\begin{center}
\includegraphics[width=85mm, clip]{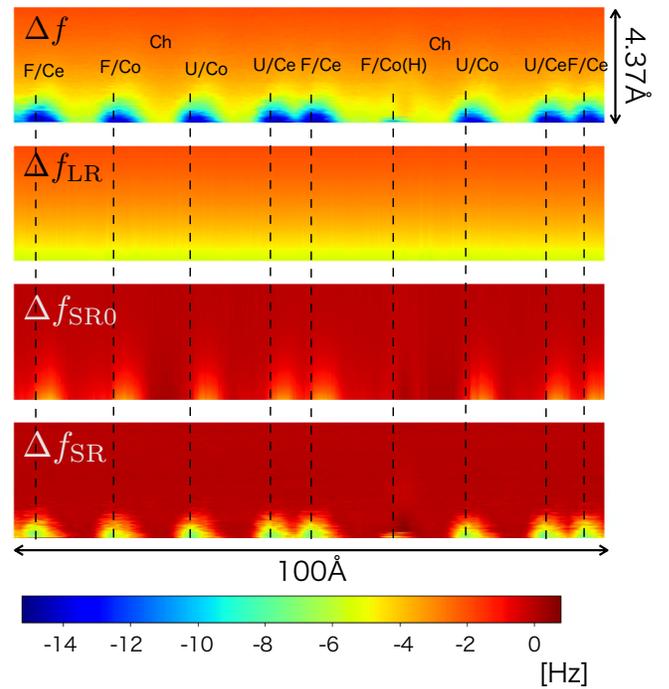}
\caption{\label{fig:SiMapLR}
Maps extracted for the case of Tip 2 using our machine-learning algorithm, and separation of the site-specific contribution apparent in the $\Delta f_{\mathrm{FIT}}$ map of Fig. \ref{fig:SiMapB}. The $\Delta f$ maps were measured over the main diagonal of the $\mathrm{Si}(111)-(7\times7)$ surface. In order to obtain the long-range interaction ($\Delta f_{\mathrm{LR}}$) without any contribution from the short-range one, we skipped the first stage of the algorithm by selecting $z_0=7.47\mathrm{\AA}$ for all sites, and computed the second stage of the algorithm fixing $d =1.5$ (see text for details). This procedure enabled us to discern a faint structure in the short-range interaction ($\Delta f_{\mathrm{SR0}}$) that is slightly sifted from the signal producing atomic contrast in the AFM images ($\Delta f_{\mathrm{SR}}$). Symbols in the maps corresponds to:“F" for faulted, “U" for unfaulted, “Ce" for center adatom, “Co" for corner adatom, “Ch" for corner hole, and “F/Co(H)" for a hydrogen atom saturating the dangling bond at a faulted corner adatom\cite{PhysRevB.87.155403}. 
}
\end{center}
\end{figure}

We can further separate the contributions in the $\Delta f_{\mathrm{FIT}}$ map acquired with Tip 2 by performing again the calculation skipping the first stage and running the second stage with a predetermined and fixed value of $z_0\ (= 7.47\mathrm{\AA}$) for all sites. This value is chosen because at such distance from the closest approach to the surface the contribution from the short-range interaction can be considered negligible.
Here, we also fixed $d = 1.5$ to obtain the other fitting parameters, which is also a reasonable value for a far enough probe-surface separation\cite{dvalue}.
In the histograms shown in Fig. \ref{fig:histogram}, the distribution of the new fitting parameters (in red) are compared with the previous ones (in grey). For completeness, we have also included the distribution of parameters for Tip 1 (in purple).
The broad distribution of the grey histograms indicates that both non-site-specific long range forces ($\Delta f_{\mathrm{LR}}$) and site-specific short-range forces ($\Delta f_{\mathrm{SR0}}$) are included in $\Delta f_{\mathrm{FIT}}$. 
On the contrary, the sharp peaks of the red histograms indicate that there is no apparent site-specific variation of the parameters, and therefore they mostly characterize the long-range part of the interaction.
This is corroborated by the maps shown in Fig. \ref{fig:SiMapLR}, where now the $\Delta f_{\mathrm{LR}}$ map is featureless, and the $\Delta f_{\mathrm{SR0}}$ map ($\Delta f_{\mathrm{SR0}}=\Delta f_{\mathrm{FIT}}-\Delta f_{\mathrm{LR}}$) shows site-specific characteristics slightly shifted from the atomic positions displayed in $\Delta f_{\mathrm{SR}}$.
Therefore, our machine-learning algorithm has allowed us to identify two contributions to the short-range force: a faint, slightly shifted atomic pattern extending over $2.2\mathrm{\AA}$ from the closest approach point ($\Delta f_{\mathrm{SR0}}$); and the standard atomic patterns ($\Delta f_{\mathrm{SR}}$) associated with the onset of the covalent bond between the last atom of the probe and the atoms at the surface, and extending approasimatelly $1\mathrm{\AA}$ from the closest approach point to the surface.

%

The presence of two contributions to the short-range interaction and the slight shift between their signatures in the $\Delta f$ signal may arise from the interaction of the surface atoms with two close-by atoms at the probe\cite{Bechstein_2009, Welker444}. 
Another origin for $\Delta f_{\mathrm{SR0}}$ could be the polarization force that has been used to image the $\mathrm{Si(111)}-(7\times7)$ surface with scanning nonlinear dielectric microscopy\cite{Yamasue:2014fq}. 
An additional plausible explanation encompasses reversible atomic displacements of the foremost atoms of the probe as it approaches the surface\cite{Sugimoto413}. 
The exact nature of the interaction giving rise to $\Delta f_{\mathrm{SR0}}$ is out of the scope of this manuscript. Nonetheless, it is worth noticing that we were able to detect this faint component of the short-range interaction because of the powerful analytic resources that our machine-learning algorithm grants for the separation of short and long contributions to the probe-surface interaction responsible for the contrast in AFM. 

%

A similar algorithm can also be used for the automatic extraction of the short-range interaction form the force curves obtained after applying an inversion procedure to the measured frequency shift.
We have applied our machine-learning algorithm to the force maps extracted from the $\Delta f$ maps displayed in Fig. \ref{fig:SiMapB}, and confirmed that the short-range forces obtained are almost identical to the ones produced by running the inversion procedure on the $\Delta f_{\mathrm{SR}} -z$ curves generated upon applying the machine-learning algorithm directly to the $\Delta f -z$ curves.

Still, there are some improvements to make on using machine-learning algorithms to separate short- and long-range interactions in force spectroscopy measurements. 
One is the effect of the noise in the spread of the distribution of the fitting parameters. 
This spread has little effect on fitting the long-range part of the curve, yet small variations in the parameters may affect the interpolation over the short-range interaction regime, slightly influencing the quantitative values of $\Delta f_{\mathrm{SR}}$, and therefore the forces. 
We are currently working on improving our algorithm to be less susceptible to experimental noise. 
Further matters to consider are the loss function and the convergence criteria, which in our case is optimized for the curves measured over the $\mathrm{Si(111)-(7\times 7)}$ surface. 
Other force sensors using ultra-small oscillation amplitudes, curves measured on other surfaces, or in different environments than UHV may require further optimization of the loss function and convergence criteria. 
To study these effects in detail, we are in the process of testing our algorithm on a variety of experimental conditions. 
We have made publicly available the algorithm used in this work elsewhere\cite{ourprogram}.

In summary, we implemented and tested a machine-learning based extraction of the short- and long-range parts of the probe-surface interaction from frequency shift maps measured along the main diagonal of the unit cell of the $\mathrm{Si(111)-(7\times 7)}$ surface. 
In the first stage of the algorithm, an appropriate application of a loss functions and the use of Ohtsu’s method enable us to determine the probe-surface separation corresponding to the onset of the short-range interaction. In a second stage, the algorithm enables us to obtain the parameters to fit the long-range part and to extract the short-range component by subtracting the fit to the measured frequency shift curves. Our method enabled us to find a faint structure at the onset of the short-range interaction for one of the probes used in the experiments that would have been otherwise obviated using human-supervised separation strategies. We believe that the method described in this manuscript holds promise to analyze two- and three-dimensional frequency shift maps composed by a large quantity of force spectroscopy curves.

This work was supported in-part by a Grant-in-Aid for Scientific Research (19H05789 and 18K19023) from the Ministry of Education, Culture, Sports, Science and Technology of Japan (MEXT).

\vspace{5mm}
\bibliography{reference}

\begin{thebibliography}{46}
\bibitem [{los()}]{lossFunction}%
  \BibitemOpen
  \href@noop {} {}\bibinfo {note} {It is important that $E_{\mathrm{1st}}$
  contains two terms. Although $\Delta \tilde{f}_{\mathrm{FIT}}$ should be
  calculated in the range of $z>\tilde{z}_0$, it is uncertain whether
  $z>\tilde{z}_0$ is the most likely value at the initial part of the first
  stage. In the first state, we set initial value of $\tilde{z}_0$ close to
  $z_\mathrm{MAX}$ so that the second term dominates initially. This enable the
  first stage calculation to finish quickly. $w_0(z)$ has the effect of
  bringing $z_0$ close to the sample surface.}\BibitemShut {Stop}%
\end{thebibliography}

\end{document}